\documentclass[twocolumn,showpacs,preprintnumbers,amsmath,amssymb]{revtex4}

\usepackage{graphicx}
\usepackage{dcolumn}
\usepackage{bm}
\usepackage{float}
\usepackage{amsmath,amssymb}

\begin{document}

\preprint{APS/123-QED}

\title{Electromagnetically induced transparency resonances inverted in magnetic field }

\author{A. Sargsyan}
\author{ D. Sarkisyan}%
 \email{davsark@yahoo.com, david@ipr.sci.am}
\affiliation{%
Institute for Physical Research, NAS of Armenia - Ashtarak-2, 0203, Armenia
}%

\author{Y. Pashayan-Leroy, C. Leroy}
\affiliation{Laboratoire Interdisciplinaire Carnot de Bourgogne, UMR CNRS 6303, University de Bourgogne, France
}%

\author{S. Cartaleva}
\affiliation{Institute of Electronics, Bulgarian Academy of Sciences, Sofia, Bulgaria
}%

\author{A.D. Wilson-Gordon}
\affiliation{Department of Chemistry, Bar-Ilan University, Ramat Gan 5290002, Israel
}%

\author{M. Auzinsh}
\affiliation{Department of Physics, University of Latvia, 19 Rainis Blvd., Riga LV-1586, Latvia
}%

\date{\today}

\begin{abstract}
The electromagnetically induced transparency (EIT) phenomenon has been investigated in a $\Lambda$-system of the $^{87}$Rb D$_1$ line in an external transverse magnetic field. Two spectroscopic cells having strongly different values of the relaxation rates $\gamma_{rel}$ are used: a Rb cell with antirelaxation coating ($L\sim$1 cm) and a Rb nanometric-thin cell (nano-cell) with thickness of the atomic vapor column $L$=795nm. For the EIT in the nano-cell, we have the usual EIT resonances characterized by a reduction in the absorption (i.e. dark resonance (DR)), whereas for the EIT in the Rb cell with an antirelaxation coating, the resonances demonstrate an increase in the absorption (i.e. bright resonances). We suppose that such unusual behavior of the EIT resonances (i.e. the reversal of the sign from DR to BR) is caused by the influence of alignment process. The influence of alignment strongly depends on the configuration of the coupling and probe frequencies as well as on the configuration of the magnetic field.
\end{abstract}

\pacs{Valid PACS appear here}
\maketitle

\section{\label{sec:level1}Introduction}

In order to observe coherent processes in atomic vapor cells, such as coherent population trapping (CPT), electromagnetically induced transparency (EIT), and electromagnetically induced absorption (EIA) it is important to realize conditions that allow ground-state coherent spin states to survive many collisions with the cell walls [1-7]. For this purpose, antirelaxation coatings (ARC) can be used in atomic vapor cells [8]. The use of a vapor cell with ARC allows one to obtain ultranarrow resonances of less than 2 Hz line-width in the nonlinear magneto-optic effect [9]. Ultranarrow EIT-resonances of width $\sim$ 100 Hz have been observed in [10] using a vapor cell with ARC. In [11], the transformation of a Ramsey EIA into a magnetic-field induced transparency resonance in a paraffin-coated Rb vapor cell in the Hanle configuration, with a line-width of 0.6 mG was reported. The Hanle configuration has also been used to study the  $^{87}$Rb, D$_1$ line [12]. The EIT process in multi-Zeeman-sublevel atoms has been considered in [13]. Experimental evidence for bright magneto-optical resonance sign reversal in Cs atoms confined in a nano-cell has been presented in [14]. Margalit $\emph{et al.}$ have studied the effect of a transverse magnetic field on the absorption spectra of degenerate two-level systems in the D$_2$ line of $^{87}$Rb [15] and degenerate $\Lambda$ systems in the D$_1$ line of $^{87}$Rb [16].

Here, we present for the first time (to our knowledge) experimental evidence for the transformation of a dark resonance into a bright resonance in an external transverse magnetic field in a Rb vapor cell with ARC. The results are compared with those obtained using a Rb nano-cell with vapor column of thickness $L$=795nm, where the sign reversal is absent.

\subsection{\label{sec:level1}Experiment}

The experimental arrangement is sketched in Fig. 1. Two beams of single-frequency extended cavity diode lasers (ECDL) with $\lambda \approx$ 795 nm (the line-width is $\sim$ 1 MHz) are well superposed and directed with the help of the polarization beam splitter PBS1 either onto the Rb ARC cell, or onto the Rb nano-cell (NC).  The coupling and probe beams have linear and perpendicular polarizations. $\emph{(1)}$-indicates Faraday isolators. The Rb ARC cell or Rb (NC) $\emph{(4)}$, are placed inside the three pairs of mutually perpendicular Helmholtz coils $\emph{(2)}$, making it possibility to cancel the laboratory magnetic field as well as to apply a homogeneous magnetic field. The optical radiations are recorded by the photodiodes $\emph{(3)}$ and the signal of the photodiodes is intensified and recorded by digital storage oscilloscope Tektronix TDS 3032B, F-are filters. The power of the coupling and probe lasers are 1- 30 mW and 5$\mu$W -0.5 mW, respectively. An improved dicroic atomic vapor laser lock (DAVLL) method is used for the coupling frequency stabilization [17]. The frequency reference spectra formation is realized with the help of an auxiliary NC of $L = \lambda$ $\emph{(5)}$ [17]. A PBS2 is used to separate the coupling and probe beams,so that only the probe beam transmission is monitored.

\begin{figure}
\includegraphics[width=8cm]{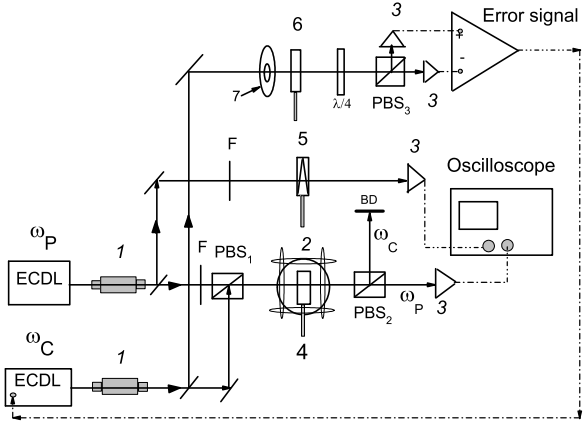}
\caption{\label{fig:1} Sketch of the experimental setup. $\emph{1}$-Faraday isolators; PBS$_{1,2,3}$ - polarization beam splitters, $\emph{2}$-Helmholtz coils; $\emph{3}$-photodiodes; $\emph{4}$-Rb vapor cell with ARC or Rb nano-cell; $\emph{5}$-auxiliary NC of $L = \lambda$; $\emph{6}$-auxiliary NC of $L = \lambda$/2; $\emph{7}$-permanent ring magnet; BD-beam damp.}
\end{figure}

\subsubsection{\label{sec:level3}The coupling is in resonance with the 1$\longrightarrow$1$^\prime$ transition; the probe is scanned through the 2$\longrightarrow$1$^\prime$ transition.}

In Fig. 2 (a), a $\Lambda$-system formed in the $^{87}$Rb D$_1$ line is shown. The coupling frequency is in resonance with the 1$\longrightarrow$1$^\prime$ transition, while the probe frequency is scanned through the 2$\longrightarrow$1$^\prime$ transition (the primes indicate the upper levels of the atoms). Two cells filled with natural Rb are used: an 8 mm-long cell with ARC (ARC cell) and a nano-cell (NC) with thickness $L$ = $\lambda$ = 795 nm [17]. An external magnetic \textbf{B}-field is directed along the probe \textbf{E$_P$} field, while \textbf{E$_C$} is perpendicular to the \textbf{B}-field (see Fig. 2 (b)). In Fig. 3, the EIT spectra for the NC (upper curve) and ARC cell (the lower curve), for a  transverse magnetic field of \textbf{B}$\approx$27 G,  are shown. The coupling and probe laser powers are 13.6 mW and 0.2 mW, respectively. It can be seen that we have the usual EIT resonances for the NC case, demonstrating a reduction in the absorption (dark resonances). As shown previously [16], there are four DRs. However, it is easy to show that the middle two must be stronger (see Figs. 19 and 20). The situation is very different for the ARC cell.

\begin{figure}
\includegraphics[width=9cm]{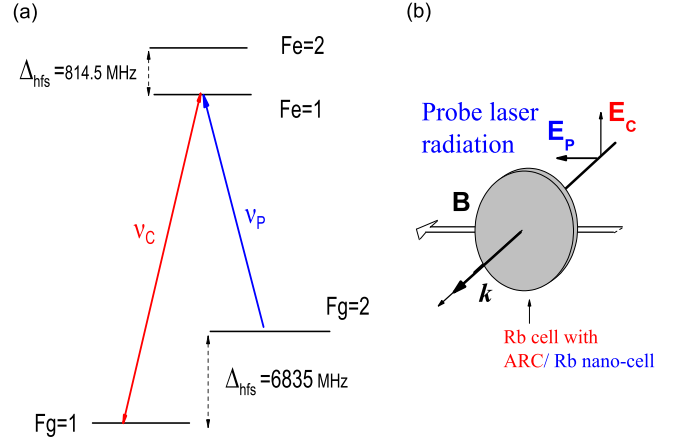}
\caption{\label{fig:2} (a) $\Lambda$-system of the $^{87}$Rb D$_1$  line, (b) the configuration of \textbf{B, k, E$_p$} and \textbf{E$_c$}.}
\end{figure}

\begin{figure}
\includegraphics[width=8.5cm]{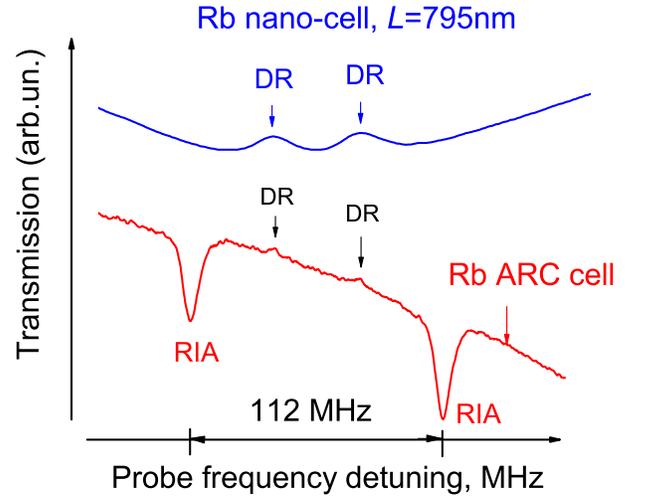}
\caption{\label{fig:3} $^{87}$Rb D$_1$ line. The EIT spectra for the Rb NC with $L$=$\lambda$=795 nm, $T$ = 110 $^\circ$C (upper curve): two DR are shown. For the ARC cell, $T$ = 36 $^\circ$C (lower curve) two small DRs and two large RIA are shown. Probe power is 0.2 mW, coupling power is 13.6 mW, $B \approx$ 27G.}
\end{figure}

There, we see that the two outer high-amplitude resonances show an increase in absorption, i.e. they are bright resonances (we call them resonances inverted by alignment (RIA)). Fig.4 shows the EIT spectra for the ARC cell as a function of the probe laser power for $B \approx$ 27 G (the coupling laser power is 13.6 mW). It can be seen that increasing the probe intensity results in the disappearance of the EIT resonance (see also Fig.5), which can be related to the stronger alignment of atomic population on Zeeman sublevels of the $F_g$ = 2 level when the probe intensity is enhanced.  The inset shows the DR for B=0, where the line-width is $\sim$5 MHz.

\begin{figure}
\includegraphics[width=8.5cm]{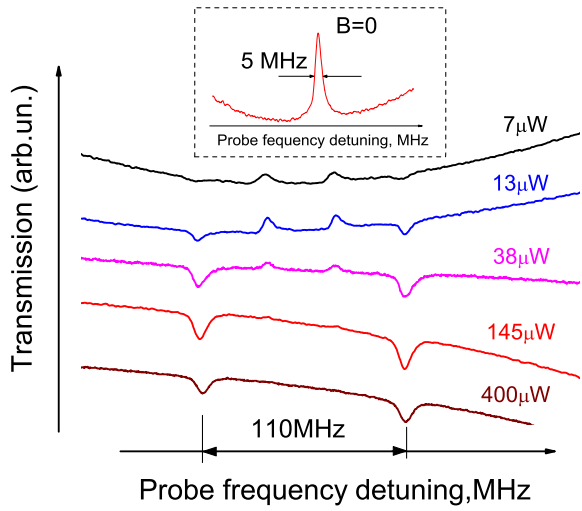}
\caption{\label{fig:4} $^{87}$Rb D$_1$ line, DR and RIA spectra for the ARC cell ($T$ = 36 $^\circ$C) versus the probe laser power. Coupling laser power is 13.6 mW, $B \sim$ 27 G.}
\end{figure}

\begin{figure}
\includegraphics[width=7cm]{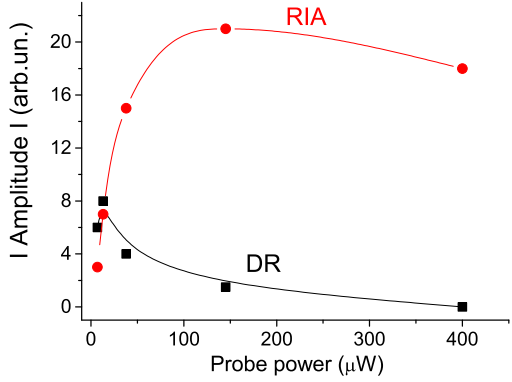}
\caption{\label{fig:5} Dependence of the amplitudes of RIA and DR as a function of the probe laser power, $T$ = 36 $^\circ$C, $B \sim$ 27G.}
\end{figure}

\begin{figure}
\includegraphics[width=8.5cm]{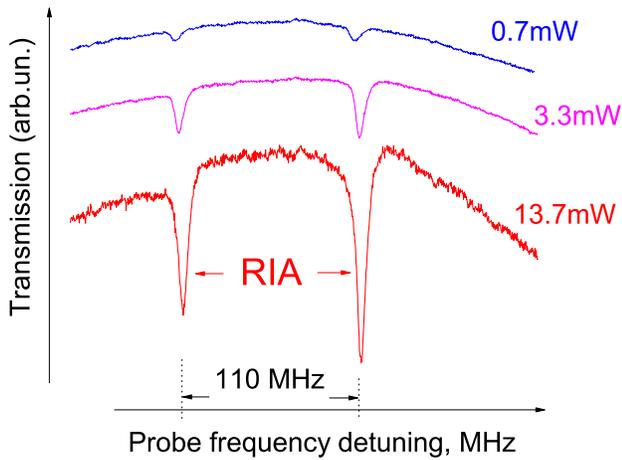}
\caption{\label{fig:6} $^{87}$Rb D$_1$ line, RIA spectra versus coupling laser power. Probe laser  power is 0.5 mW, $B \sim$  27 G.}
\end{figure}

 Note, that for the coupling and probe lasers powers 1mW and 0.01 mW, respectively, the line-width of DR is 3.5 MHz (but with a smaller contrast). Is it well-known that in order to get narrow DRs, the two lasers must be coherently coupled [3,4]. However, for the study of the EIT resonance splitting and shift in strong external magnetic fields (when the frequency shift reaches several GHz) a convenient alternative is the use of two different lasers [19]. The dependence of the DR and RIA for the ARC cell as a function of the coupling power (the probe power is 0.5 mW), is shown in Figs.6 and 7. The enhancement of the amplitude of the RIA resonances with increasing coupling beam intensity can be attributed to hyperfine optical pumping from the $F_g$=1 to $F_g$=2 level due to the coupling laser.  As a result, more atoms are redistributed among the $F_g$=2   Zeeman sublevels due to the Zeeman optical pumping (ZOP) process. In order to explain the unusual behavior of the RIA, the transition system in terms of ZOP for the $^{87}$Rb D$_1$ line 2$\longrightarrow$1$^\prime$ is shown in Fig. 8. Due to the ZOP effect caused by the strong probe beam, the population of level $F_g$ = 2 can be trapped in the sublevels with $m_F$ = $\pm$2 (shown by black bars), i.e. so-called alignment occurs [20,21]. The alignment effect causes a strong redistribution of the population of the $F_g$=2 level. Note that the population of the outermost Zeeman sublevels with $m_F$ = $\pm$2 is high also due to the fact that these levels are not excited by the laser light.
\begin{figure}
\includegraphics[width=7cm]{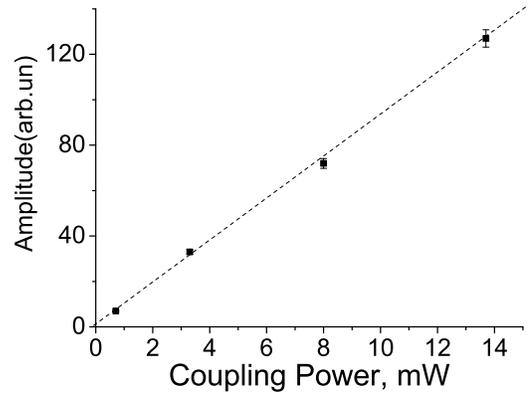}
\caption{\label{fig:7} Dependence of the RIA amplitude versus coupling laser power, probe laser power is 0.5 mW, $B \approx$ 27 G.}
\end{figure}

 \begin{figure}
\includegraphics[width=7.5 cm]{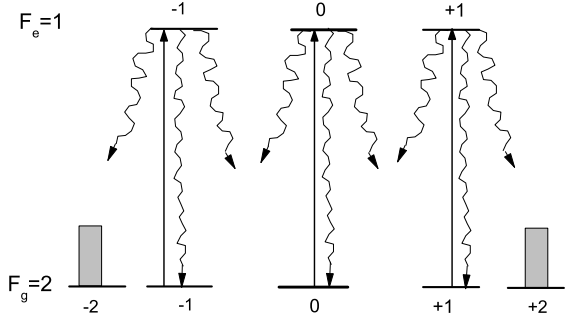}
\caption{\label{fig:8} $^{87}$Rb D$_1$ line,  Zeeman optical pumping process for the 2$\longrightarrow$1$^\prime$ transition [see 18,19].}.
\end{figure}

The increase (or reduction) in the population $N$ by $\Delta$$N$ of the levels $F_g$=2 with $m_F$ = $\pm$2 is
 \begin{eqnarray}
\Delta N \sim (\frac{dE_P} {\hbar})^2/(\gamma_N \gamma_{rel})
\label{eq:one}.
\end{eqnarray}
[20,21], where  $(dE_P\backslash\hbar)$ is the Rabi frequency of the probe laser, $\gamma_N$ is the natural linewidth and $\gamma_{rel}$ is the rate of decay of the coherence between the lower levels. We see from Eq. (1) that $\Delta N$ strongly depends on the relaxation rate $\gamma_{rel}$. For a usual cell without ARC, it is inversely proportional to the time-of-flight of the atom across the laser beam with diameter $D$, and for $D\sim$1 mm, $\gamma_{rel}$ is typically $\sim$100 kHz. For our ARC cell, it is $\sim$ 1 kHz due to the ARC not being of high quality, while an advantage of this ARC  coated cell is its resistance to a temperature up to 100 $^\circ$C [22] .

 Meanwhile for the NC, the relaxation rate is much higher ($\gamma_{rel}$ is $\sim$1 MHz). This is related to the frequent collisions of the Rb atoms with the NC windows, destroying the alignment process (see the upper curve in Fig. 3). Fig. 9 shows the splitting and the shift of the RIA as a function of the transverse magnetic B field. The frequency interval between the RIAs is $\approx$ 4.2 MHz/G (see below) and remains linear with $B$ up to 100 G. As can be seen, the amplitudes of the RIA are almost independent of the magnetic field.

\begin{figure}
\includegraphics[width=9cm]{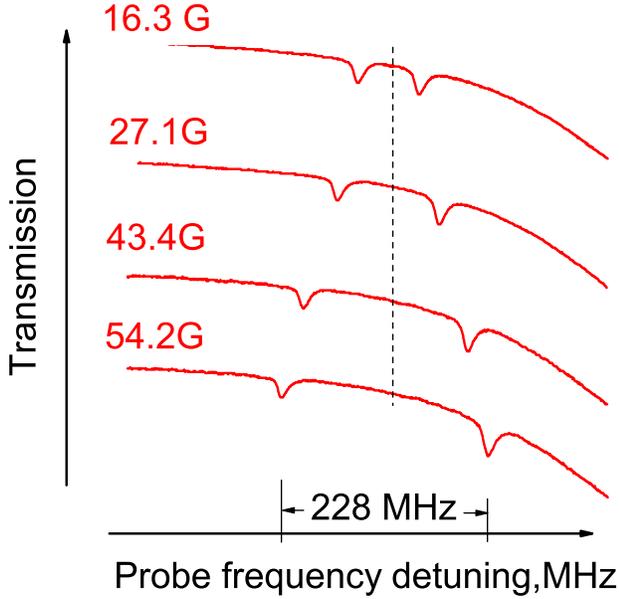}
\caption{\label{fig:9} Splitting and shift of the RIA as a function of the external transverse magnetic B-field. The frequency interval between the RIA is $\approx$ 4.2 MHz/G. The dashed line shows that RIAs are displaced symmetrically with respect to the initial position of the DR at $B$=0.}
\end{figure}
As to the physical mechanism of RIA formation we propose the following. Since there are eight Zeeman sublevels for the $F_g$ = 1, 2 levels, initially all the Zeeman sublevels will be equally populated with population 1/8 for each sublevel. In the case of strong ZOP, all the population of $F_g$=2 is concentrated in the sublevels $m_F$ = $\pm$2, so that their population will be 5/16, while the population in the $F_g$=1 Zeeman sublevels is 2/16 for each sublevel. Since the population $N(F_g=2, m_F = \pm2) > N(F_g =1, m_F = 0,\pm1)$, strong absorption of the probe radiation $\nu_P$ can occur via the two-photon Raman-type process (with emission at the coupling frequency $\nu_C$), shown in Fig. 10. We suppose that the sublevels $m_F=\pm2$ of the level $F_e$=2 are also involved (via the 2$\longrightarrow$2$^\prime$ transition) in this two-photon Raman-type process (see Fig. 11). Note, that competition between the EIT and a two-photon Raman-type process has been observed for a $\Lambda$-system of Na atoms when the probe and coupling laser frequencies are near resonance [23].
\begin{figure}
\includegraphics[width=8cm]{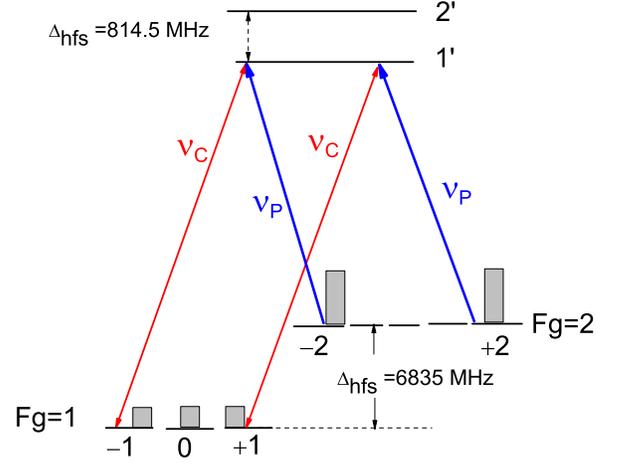}
\caption{\label{fig:10} $^{87}$Rb D$_1$ line, two-photon Raman-type process. The populations of $N(F_g = 2,m_F = \pm2)$ are indicated by the large bars, and the populations $N(F_g = 1,m_F = \pm1)$) by the small bars. We suppose that the sublevels $m_F = \pm2$ of the level $F_e$=2 are also involved (via the 2$\longrightarrow$2$^\prime$ transition).}
\end{figure}

A similar process has been considered in a $\Lambda$-system of $^{87}$Rb, where the ground levels of the system are formed by the Zeeman sublevels of the same ground level [24]. The cross-section for such two-photon Raman-type absorption (TPA) process of the probe beam $\nu_P$ is

\begin{eqnarray}
\sigma_{TPA}=\frac{\lambda^2} {16\pi^2} \frac{\gamma_N} {\gamma_{21}}(\frac{dE_C} {\hbar\Delta})^2
\label{eq:two}.
\end{eqnarray}
where $(dE_C/\hbar)$ is the Rabi frequency of the coupling radiation ($\sigma_{TPA}$ is thus proportional to the coupling laser power, see Figs.6 and 7),  $\Delta$ is the laser frequency detuning from the 5P$_{1/2}$ level, $\gamma_N$ = 6 MHz, $\gamma_{rel} \sim$ 0.001 MHz, $\Lambda$ = 795 nm. For TPA of the probe laser radiation one has $\sigma_{TPA} (N_{g2}  -N_{g1}) L, \Delta \sim30-40$ MHz , $\Omega_C \sim$ 35 MHz, $L \sim$ 0.8 cm. If we suppose that $^{87}$Rb ($N_{g2} - N_{g1}) \sim 10^8 cm^{-3}$, (this is realistic number since N ($^{87}$Rb) $\sim7\times10^9 cm^{-3}$ for $T=35 ^\circ$C), then TPA (i.e. RIA) can reach $\sim$ 10 percent, which agrees with the experiment.

\begin{figure}
\includegraphics[width=9cm]{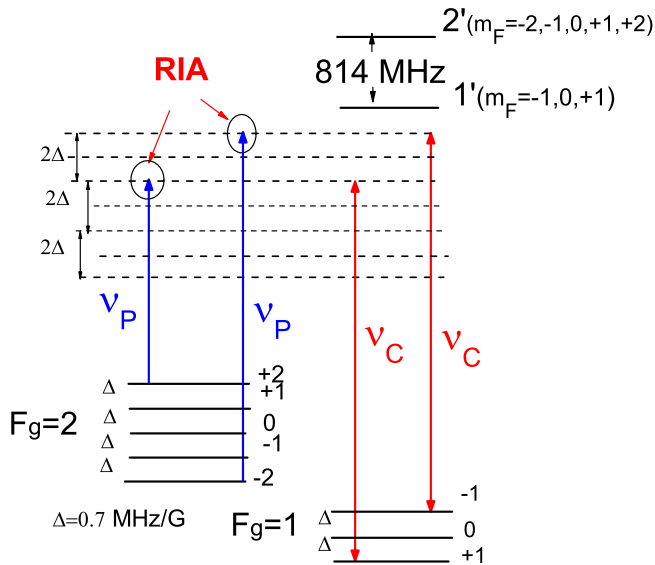}
\caption{\label{fig:11} $^{87}$Rb D$_1$ line, two-photon Raman-type process. The frequency separation between the RIAs: the coupling laser (with $\sigma^+$ and $\sigma^-$ polarizations) frequency $\nu_C$ is fixed, the probe laser (of $\pi$-polarization) frequency $\nu_P$ is varied. The frequency interval between RIAs is 6$\times$0.7MHz/G= 4.2 MHz/G.}
\end{figure}

The value of the frequency separation between the RIAs is determined by the diagram presented in Fig. 11. The frequency $\nu_C$ of the coupling laser (having $\sigma^+$ and $\sigma^-$ polarizations) is fixed, while the frequency $\nu_P$ of the probe laser (having $\pi$ polarization) varies. The first process forming RIA starts from $F_g$ = 2, $m_F$ = +2 and finishes in $F_g$ =1, $m_F$ = +1, while the second process starts from $F_g$ = 2, $m_F$ = -2 and finishes in $F_g$ = 1, $m_F$ = -1. We suppose that the sublevels $m_F$ = $\pm$2 of the level $F_e$=2 are also involved for the probe radiation (via the 2-2$^\prime$ transition) by this two-photon Raman-type process, since level 2$^\prime$ is located at $\sim$800 MHz above level 1$^\prime$, so that the transitions 2-1$^\prime$ and 2-2$^\prime$ are overlapped by the Doppler wings. Note, that it has been demonstrated  [25] for a TPA process (in a ladder-system) that, even for 1.5 GHz-detuning from the intermediate level, nearly 100 percent absorption is achieved for a Rb 2mm-long cell. A similar situation takes place for the two-photon Raman-type process shown in Fig.15 (i.e. the sublevels $m_F$ = $\pm$2 of the level $F_e$=2 are also involved).
	As we see, the frequency interval between two RIAs is 6$\times$0.7MHz/G= 4.2 MHz/G. It is easy to show that for the EIT splitting in the case of the NC (Fig. 3), the frequency interval between the two neighboring EIT resonances is 2$\times$0.7MHz/G= 1.4 MHz/G.

\subsubsection{\label{sec:level3}The coupling is in resonance with the 2$\longrightarrow$1$^\prime$ transition; the probe is scanned through the 1$\longrightarrow$1$^\prime$ transition.}

\begin{figure}[H]
\includegraphics[width=5cm]{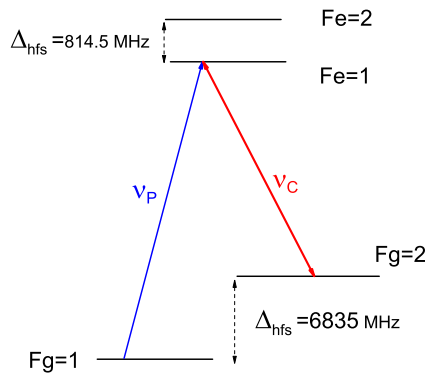}
\caption{\label{fig:12} $\Lambda$-system of the $^{87}$Rb D$_1$ line.}
\end{figure}

\begin{figure}[H]
\includegraphics[width=6cm]{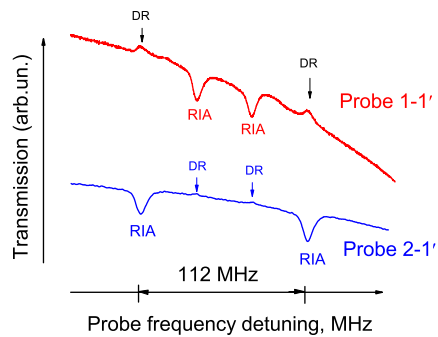}
\caption{\label{fig:13} $^{87}$Rb D$_1$ line,the EIT spectra for the ARC cell, $T$= 36 $^\circ$C. The upper curve shows the spectrum for the coupling and probe laser frequency configuration of Fig. 12, while the lower curve refers to the configuration of Fig. 3. Probe power is 0.2 mW, coupling power is 13.6 mW, $B \approx$ 27G. }
\end{figure}

Fig. 13 presents the transmission spectra for the ARC cell, at T= 36 $^\circ$C. The upper curve shows the spectrum for the coupling and probe laser frequency configuration presented in Fig. 12, while the lower curve corresponds to the configuration of $\nu_C$ and $\nu_P$ presented in Fig.3. It is important to note that when the probe power is increased, DRs (shown in the upper curve) are still present in the spectrum, while in the lower curve the DRs completely vanish for power $>$0.4 mW (only the two RIAs remain). Thus, we suppose that the alignment process illustrated in Fig.14 does not lead to complete depletion of the $m_F$ = $\pm$1 Zeeman sublevels of the ground state. Probably this can be related to the hyperfine optical pumping by the coupling laser to the $F_g$ =1 level. Due to the ZOP effect caused by the probe laser in the 1$\longrightarrow$1$^\prime$ Zeeman sub-level system, the main population of the level $F_g$ =1 in the case of strong probe laser is concentrated in the sublevels $m_F$ = 0 (shown by black bar) [18,19]. The value of the frequency separation between the RIAs is determined by the diagram presented in Fig. 15. The coupling laser frequency $\nu_C$ is fixed, while the probe laser frequency $\nu_P$ is varied and has $\pi$ polarization. The first process forming RIA starts from $F_g$ =1, $m_F$ = 0 and finishes in $F_g$ = 2, $m_F$= +1, while the second process starts from $F_g$ = 1, $m_F$ = 0 and finishes in $F_g$ = 2, $m_F$ = -1.  Note that the RIA resonance formation starts from the most populated Zeeman sublevel of the $F_g$ = 1 hyperfine level which is $m_F$ = 0 (see Fig. 14). Thus both RIA resonances (Fig. 13, upper spectrum) are situated in the middle, between the DRs. For good DR formation, the atomic population of ground levels in the $\Lambda$-system has to be similar. By contrast, the RIA resonance is well pronounced when there is a large population difference between the ground levels. Thus, the spectrum presented in Fig. 13 (the lower spectrum) can be explained by the fact that, here, the most populated Zeeman sublevels of the $F_g$ = 2 level are $m_F$ = $\pm$2. Consequently, the DRs are between the RIA resonances. As we see the frequency interval between two RIAs (the upper spectrum) is 2$\times$0.7MHz/G= 1.4 MHz/G.

\begin{figure}
\includegraphics[width=4.5cm]{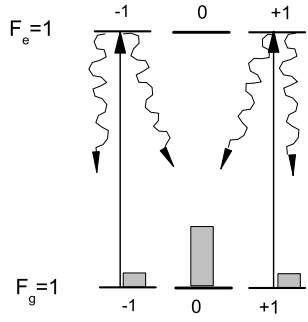}
\caption{\label{fig:14} $^{87}$Rb D$_1$ line, , Zeeman optical pumping process for
 the 1$\longrightarrow$1$^\prime$ transition [18,19].
}
\end{figure}

\begin{figure}
\includegraphics[width=7cm]{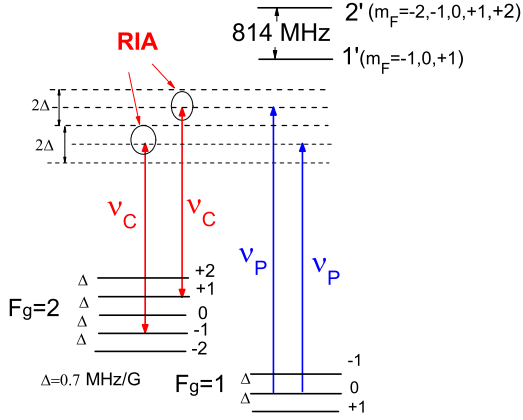}
\caption{\label{fig:15} $^{87}$Rb D$_1$ line, two-photon Raman-type process. The coupling laser contains $\sigma^+$ and $\sigma^-$ polarizations and its frequency $\nu_C$ is fixed, the probe laser is of the frequency $\nu_P$ that is varied and has $\pi$-polarization. As we see the frequency interval between RIAs for the upper curve presented in Fig.13 is 1.4 MHz/G. }
\end{figure}

\subsubsection{\label{sec:level3}The coupling is in resonance with the 1$\longrightarrow$2$^\prime$ transition; the probe is scanned through the 2$\longrightarrow$2$^\prime$ transition.}

\begin{figure}[H]
\includegraphics[width=5cm]{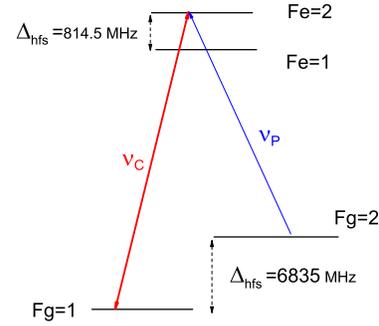}
\caption{\label{fig:16} $\Lambda$-system of the $^{87}$Rb D$_1$ line. }
\end{figure}

\begin{figure}[H]
\includegraphics[width=6cm]{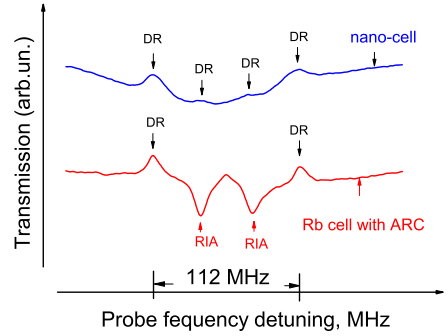}
\caption{\label{fig:17} $^{87}$Rb D$_1$ line, the EIT spectra for the Rb nano-cell, $L=\lambda$=795 nm, T= 110 $^\circ$C (upper curve) and for the Rb ARC cell, T= 36 $^\circ$C (lower curve), the probe laser power 0.2 mW , the coupling laser power is 13.6 mW, $B \approx$ 27G.  }
\end{figure}

The $\Lambda$$^{87}$Rb system of D$_1$ line is shown in Fig. 16. Fig. 17 shows the EIT and RIA spectra for the NC, $L$= $\lambda$ =795 nm (the upper curve) and the ARC cell (the lower curve), for the transverse magnetic field $B \approx$ 27 G. The coupling and probe laser powers are 13.6 mW and 0.2 mW, respectively. As can be seen for the EIT spectrum in the NC, all four narrow features represent usual DRs characterized by a reduction in the absorption. Under similar conditions in the case of the ARC cell, the two DRs are of reversed sign, i.e. RIA resonances that show increased absorption. The frequency interval between the two RIAs in the lower curve of Fig. 17 is 2$\times$0.7MHz/G= 1.4 MHz/G. Here, the  ZOP results in strong population accumulation in the $F_g$=2, $m_F$=0 Zeeman sublevel. Thus, the situation is similar to that presented in Figs. 13 and 14, and the RIA resonances are situated between the DRs.

\subsubsection{\label{sec:level3}The coupling is in resonance with the 2$\longrightarrow$2$^\prime$ transition; the probe is scanned through the 1$\longrightarrow$2$^\prime$ transition.}

\begin{figure}[H]
\includegraphics[width=5cm]{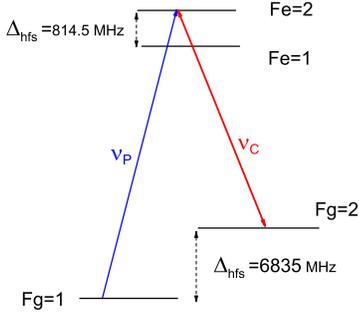}
\caption{\label{fig:18} $\Lambda$-system of the $^{87}$Rb D$_1$ line. }
\end{figure}

\begin{figure}[H]
\includegraphics[width=6cm]{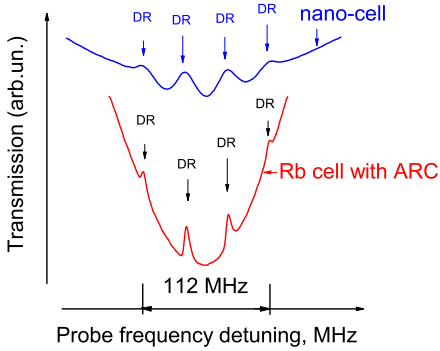}
\caption{\label{fig:19} $^{87}$Rb D$_1$ line, the EIT spectra for the NC, $L=\lambda$=795 nm, T= 110 $^\circ$C (upper curve) and for the ARC cell, T= 36 $^\circ$C (lower curve).Probe laser power is 0.2 mW and, coupling laser power is 13.6 mW, $B \approx$ 27G.  }
\end{figure}

This case is an interesting one because it differs from the previously presented results. Fig. 19 (the upper curve) shows that there are only EIT resonances in the spectrum, for the case of the NC with  $L$= $\lambda$=795nm. All DRs have good amplitudes that can be related to the small value of Zeeman sublevel population redistribution due to the alignment process. In Fig. 19 (lower curve) the case of using the ARC cell at the transverse magnetic field of $B \approx$ 27 G is presented, for the configuration shown in Fig. 18 and coupling and probe laser powers of 13.6 mW and 0.2 mW, respectively. As we can see, it is only for this particular configuration that the EIT spectra, both for the NC and the ARC cell, have the usual EIT resonances characterized by a reduction in absorption. The frequency interval between the neighboring EITs (see Fig. 20) is 2$\times$0.7MHz/G =1.4 MHz/G. For this energy level configuration and magnetic field, the transitions between Zeeman sublevel of the $F_g$=1$\longleftrightarrow$$F_e$=2 hyperfine transition will produce alignment of atomic population of Zeeman sublevels belonging to the $F_g$=1 level. However, in this configuration all Zeeman sublevels of the $F_g$=1 level are excited by the probe light beam. Thus, there is no possibility to achieve strong ZOP thereby creating conditions for RIA resonance formation.

\begin{figure}[H]
\includegraphics[width=8cm]{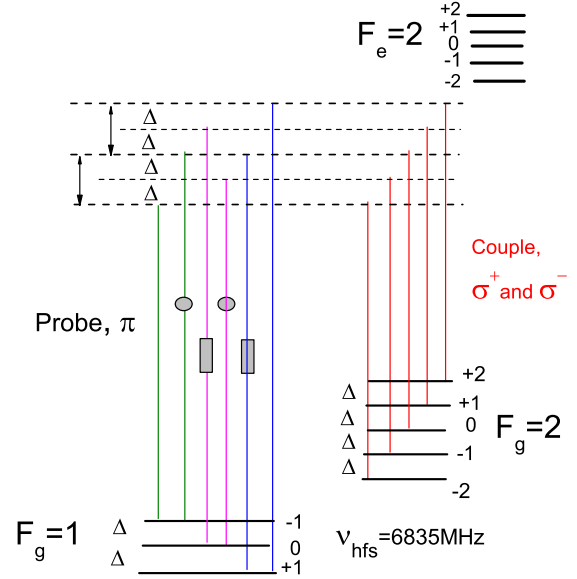}
\caption{\label{fig:20} $^{87}$Rb D$_1$ line, two-photon Raman-type process. The coupling laser has the fixed frequency $\nu_C$ and $\sigma^+$ and $\sigma^-$ polarizations, the probe laser has $\pi$-polarization and the frequency $\nu_P$ is varied. Degenerate frequencies are indicated by the circles and rectangles. Thus, there are four different EIT frequencies. The frequency interval between the neighboring EITs is 2$\times$0.7MHz/G= 1.4 MHz/G. }
\end{figure}

	Note, that in [18] the results for EIT in magnetic field are presented for the $^{87}$Rb D$_1$ line for the $\nu_C$ and $\nu_P$ configuration of Fig.18, and that there are also four EIT resonances, as shown in Fig.19. However, the diagram shown in Fig. 1 of [18] is incorrect, because the coupling laser frequency varies, rather than being constant as shown in Fig. 20.

\subsection{Discussions and conclusion}

As we see from the results presented above, the most convenient $\Lambda$-system for RIA observation is the one shown in Fig. 2 (a), where $\nu_C$ is in resonance with the 1$\longrightarrow$1$^\prime$ transition, while $\nu_P$  is scanned through the 2$\longrightarrow$1$^\prime$ transition. We suppose that for this case (the configuration shown in Fig. 2(b)), the alignment is very effective due to strong Zeeman optical pumping. For the other $\nu_C$  and $\nu_P$  configurations the alignment is weaker, which is probably caused by the additional redistribution of the population between the Zeeman sublevels caused by the strong coupling laser.

In order to demonstrate that the alignment process (shown in Fig. 6) is caused by the probe laser with $\pi$ polarization, we have studied another experimental configuration shown in Fig. 21. The EIT spectrum for the ARC cell and magnetic field $B \approx$ 27 G is shown in the lower spectrum of Fig. 22, while the upper spectrum shows the DRs for $B=$0.  As we can see, the spectrum is very different from that presented in Fig. 3, i.e. instead of the RIAs we have four DRs characterized by a reduction in the absorption (the two middle DRs have dispersion-like profiles that probably demonstrate the competition between the EIT and TPA).

\begin{figure}[H]
\includegraphics[width=5cm]{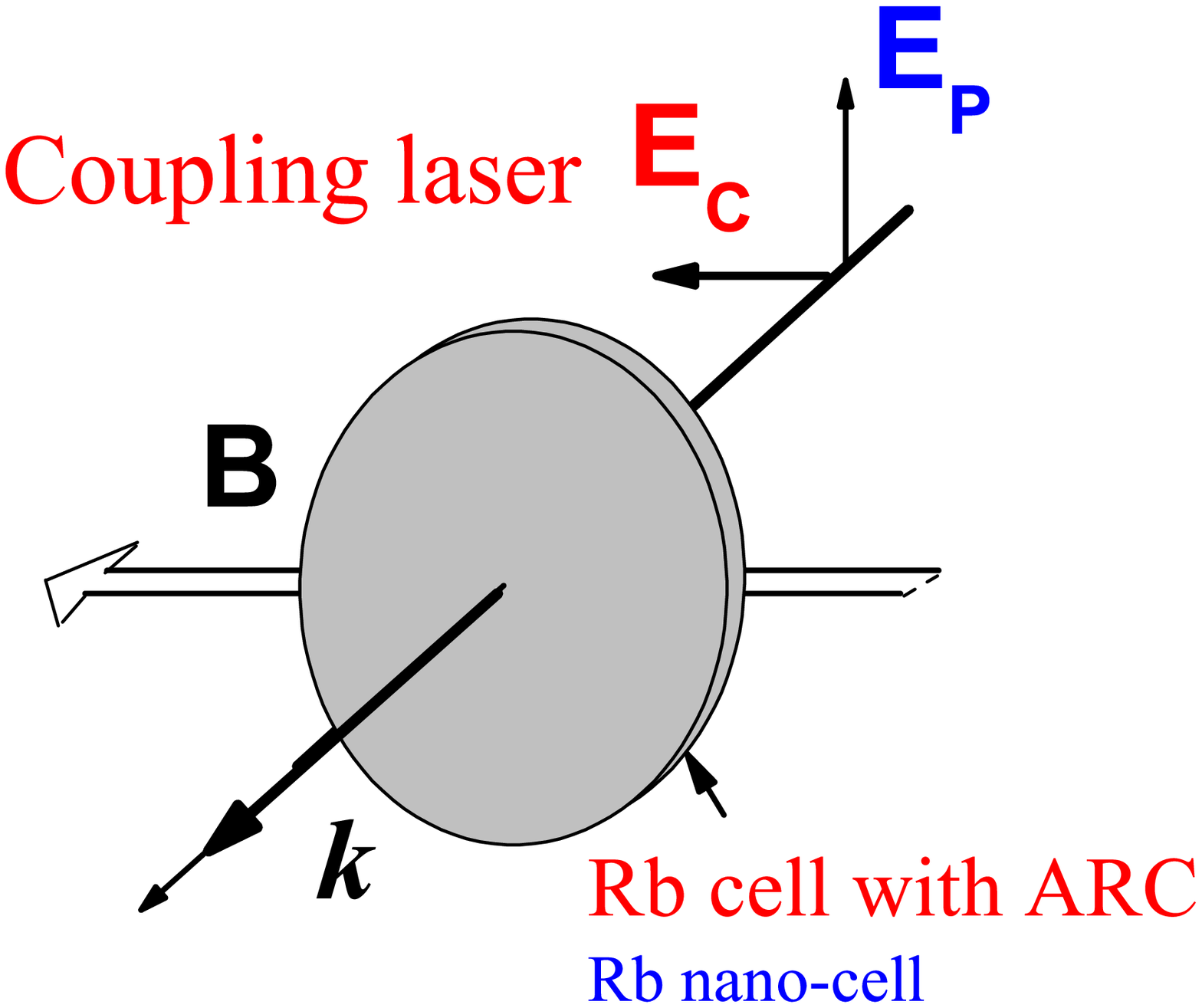}
\caption{\label{fig:21} The configuration of \textbf{B, k, E$_P$} and \textbf{E$_C$}.}
\end{figure}

\begin{figure}[H]
\includegraphics[width=7cm]{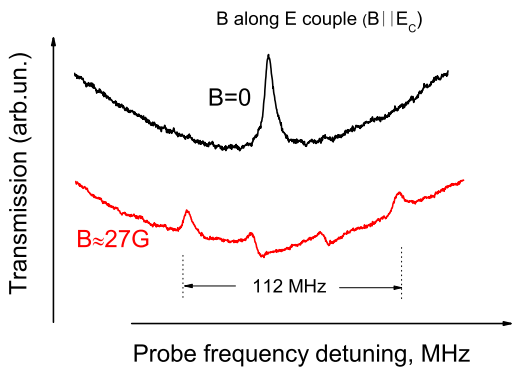}
\caption{\label{fig:22} $^{87}$Rb D$_1$ line, the EIT spectra for the ARC cell, T= 36 $^\circ$C. The coupling and probe laser powers are 13.6 mW and 1 mW, respectively. The upper spectrum shows the EIT resonance for $B$=0, the lower spectrum shows the EIT spectrum for  \textbf{B}$\parallel$\textbf{E}$_C$, $B \approx$ 27 G .  }
\end{figure}

Also, using the $\nu_C$  and the $\nu_P$  configuration shown in Fig.2 (a), and applying a longitudinal magnetic field (\textbf{B} $\parallel$ \textbf{k}) we have detected the well-known spectrum with three DRs shown in Fig. 23. In this case the magnetic field creates three $\Lambda$-systems (shown in the inset) and three DRs are formed [3,16,26]. The middle DR has larger amplitude, since there are two channels for its formation when the probe and coupling lasers have polarizations ($\sigma^+$, $\sigma^-$) or ($\sigma^-$, $\sigma^+$) and they form the middle DR with the same frequency. The identical three DRs are formed also in the NC with $L$ = 795 nm [17].

\begin{figure}[H]
\includegraphics[width=7cm]{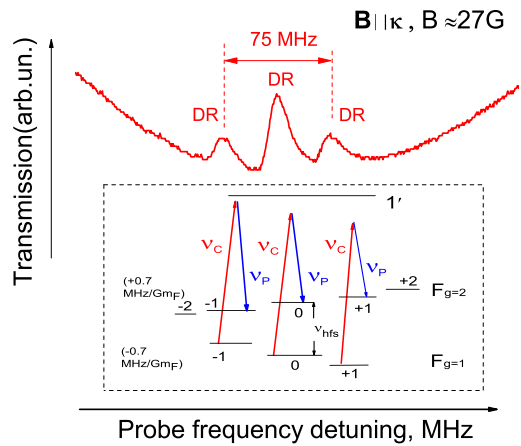}
\caption{\label{fig:23} $^{87}$Rb D$_1$ line, the EIT spectra for the ARC cell, T= 36 $^\circ$C. The coupling and probe laser powers are 13.6 mW and 1 mW, respectively, and a longitudinal magnetic field \textbf{B}$\parallel$\textbf{k}, $B \approx$ 27 G is applied. Three DRs are shown. The inset shows three $\Lambda$-systems formed by the magnetic field.   }
\end{figure}

From the systematic experimental study reported here, it can be concluded that the good amplitude RIA resonances are observed in the case of strong alignment enabled by the probe beam and low Zeeman optical pumping by the coupling laser.

	It is well-known that velocity selective optical pumping (VSOP) resonances accompany the EIT (DR) resonances [27]. In the case of the EIT in an atomic vapor cell with ARC (see Figs. 3, 9, 13, 17, 19, 22 and Fig.23) these VSOPs are of very small amplitude or even absent.

	An easy practical method to measure the quality of the ARC is presented in [28], which is based on the measurement of the temporal dependence of the intensity of the fluorescence from irradiated alkali atoms by pulsed light at resonance. Note that the results presented in Fig. 4 can be also used for testing the quality of the ARC, i.e. to determine the number of the collisions N$_C$(surv) with the cell walls which is sufficient for the ground-state coherent spin states to survive. As we see for the probe power of 13 $\mu$W the ratio of the amplitudes $\mid$A(RIA)/A(DR)$\mid \approx $1. If N$_C$(surv) is larger than that used for the ARC cell (N$_C$(surv) $\approx$200-300), the ratio will be $>$1 at the same probe laser power of 13 $\mu$W.

	A theoretical model of the $\Lambda$-system (shown in Fig.2(a)) taking into account eight Zeeman sublevels of the ground levels and 3 sublevels of the excited level has been included and the calculations are in progress.

\begin{acknowledgments}
This work was partially supported by a Marie Curie International Research Staff Exchange Scheme Fellowship within the 7th European Community Framework Programme: $''$Coherent optics sensors for medical applications-COSMA$''$ (Grant Agreement No: PIRSES-GA-2012-295264) and in the scope of the International Associated Laboratory IRMAS (CNRS-France $\&$ SCS-Armenia).
\end{acknowledgments}

\end{document}